\documentclass[12pt]{article}

\usepackage{hyperref}
\usepackage{graphicx}

\begin{document}

\title{On the Metric Structure of Space-Time\thanks{first appeared in:
M. A. del Olmo, M. Santander, and J. Mateos Guilarte, eds., Group Theoretical Methods in Physics, Vol. II (Proceedings of the XIX Int. Colloquium, Salamanca, Spain, June 29 - July 4, 1992), Anales de Fisica, Monografias 1, CIEMAT, Madrid, 1993, ISBN 84-7834-160-9, pp. 483-486
}
}

\author{Jochen Rau\thanks{email: jochen.rau@q-info.org} \\ 
Institut f\"ur Theoretische Physik \\
Johann Wolfgang Goethe-Universit\"at \\
Max-von-Laue-Str. 1, 
60438 Frankfurt am Main,
Germany
}

\date{\today}

\maketitle

\begin{abstract}
I present an analysis of the physical assumptions needed
to obtain the metric structure of space-time.
For this purpose
I combine the axiomatic approach pioneered by Robb
with ideas drawn from
works on Weyl's {\em Raumproblem}.
The concept of a Lorentzian manifold is replaced
by the weaker concept of an `event manifold',
defined in terms of volume element, causal
structure and affine connection(s).
Exploiting properties of its structure
group, I show that distinguishing Lorentzian manifolds
from other classes of event manifolds requires the key idea of 
General Relativity:
namely that the manifold's physical structure, rather than being fixed,
is itself a variable.
\end{abstract}

\newpage

\section{Introduction}

In General Relativity, space-time is assumed to be
a Lorentzian manifold.
The metric field determines the volume element
and the causal structure.
Conversely, given a Lorentzian manifold, 
the causal structure and the volume element
uniquely specify the metric field.
But why does one start from
a Lorentzian manifold in the first place?
Why are light signals
described by a quadratic equation
$g_{ab}d x^ad x^b=0$?
Why is the light `cone' not
a pyramid standing on its top?
One might argue that the quadratic term is the lowest order
contribution to the Taylor expansion
of some distance function;
but it is unclear both how this distance function
is motivated and, if it is,
whether the quadratic term
is necessarily non-trivial.
Indeed there have been suggestions that rather than
being Lorentzian, the space-time geometry might
be Finslerian \cite{1}.
So why do
Lorentzian manifolds nevertheless play a privileged role?
 
This is a particular case of
Weyl's space problem.
It was first solved by Weyl \cite{2}, who also
placed it in the context of
General Relativity.
A more elegant treatment was given by
Cartan \cite{3}.
Since then their result,
known as Weyl-Cartan theorem, has been reviewed
by various authors \cite{4}.
A very different line of reasoning was initiated by Robb, who
for the case of Special Relativity 
derived the Minkowskian metric from
properties of the causal relations
{\em before} and {\em after} \cite{5}.
More recent attempts to axiomatize General Relativity
in Robb's spirit are based on such notions as
signals, light rays, freely falling particles,
or clocks \cite{6}; 
some even invoke quantum mechanics \cite{7}.
 
A synthesis of these two kinds of approaches is the aim of the present paper.
Its purpose is
not so much an actual derivation
as it is an analysis:
which physical assumptions are being tacitly made
whenever one postulates the existence of a 
Lorentzian metric?
Only after these assumptions are exhibited
can one start to systematically relax them;
thus, answers to the above question may be helpful for the
study of more general space-time structures.
 
Primitive concepts are taken to be events, counting of events,
causal relationships and the ability to compare measurements;
the corresponding mathematical structures are a
differentiable manifold, volume element, causal vectors
and affine connection(s),
leading to the notion of an
`event manifold'.
The key assumption, which I will call `deformability',
is that
the event manifold's physical structure is allowed to vary freely.
The proof of the Weyl-Cartan theorem
is then reviewed to establish the result that
any deformable
event manifold must be Lorentzian.
\begin{figure}
	\centering
		\includegraphics[width=\textwidth]{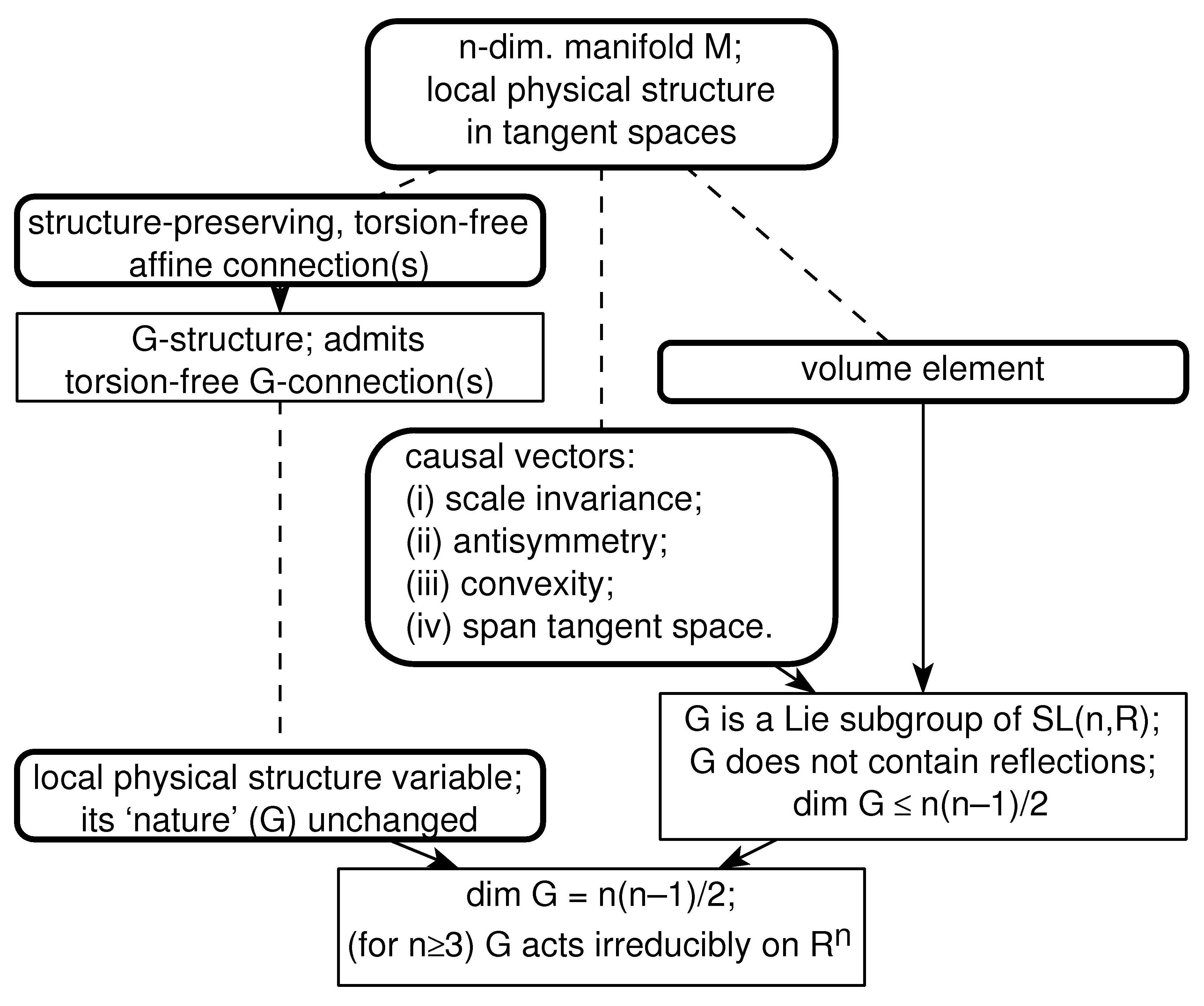}
	\caption{Line of argument.}
	\label{lofa}
\end{figure}

\section{Event Manifolds}

I assume that space-time is a
connected $n$-dimensional 
differentiable manifold $M$.
At $x\in M$, local
measurements (e.g., evaluating vector
fields) are performed using a basis of the tangent space $T_xM$.
In order to have a means to compare local measurements at
different points,
I require the manifold to be endowed with an affine 
connection. 
The connection is assumed to be torsion-free.
Space-time is also endowed with a physical structure $\Phi$
reflecting, e.g., causal relationships between events.
Details of this structure do not matter at this point;
all I assume is that (a) $\Phi$ induces in each
$T_xM$ a local physical structure $T_x\Phi$,
and (b) this local structure can be measured
with a basis of $T_xM$.
In order for an observer at $x\in M$ to be able
to determine the local structure
not only at $x$ itself but also
elsewhere on the manifold,
the connection must preserve the local physical structure.
  
A frame field provides a map
from the local physical structure
in $T_xM$ to some structure in $R^n$
 for every $x\in M$. The images in $R^n$ will generally
 vary from point to point. However,
since the connection is structure-preserving,
one can use parallel transport to construct a
special frame field called an $n$-bein 
such that the image of the local physical structure
is the same everywhere.
This enables one to fix a standard structure
$\eta$
in $R^n$ and then describe the local physical structure
by the $n$-bein.
However, there
may be several $n$-beins associated with
the same local physical structure.
Such a redundancy in the mathematical
description of a physical structure gives rise
to a gauge theory.
Its symmetry group is
the Lie subgroup $G$ of $GL(n,R)$ which leaves
$\eta$ invariant.
Hence, the local physical structure
on the manifold $M$ corresponds to an entire subbundle of
the frame bundle $F(M)$, with (reduced) structure group $G$.
This is a $G$-structure on $M$ \cite{8}.
It must admit a connection,
called a $G$-connection.
In contrast to Weyl, I do not assume the
$G$-connection to be determined uniquely.
   
A discrete rather than continuous set of events
can be characterized by
the number of events and 
their mutual causal relationships.
By analogy one expects that on a continuous space-time manifold,
the physical structure $\Phi$ should be a pair $(\mu,\le)$
consisting of a volume element
and a causal relation.
A differentiable map $\gamma: [0,1] \to M$
with the property that
$\gamma(t_1)\le \gamma(t_2)$ iff
$t_1\le t_2$ is called a causal curve.
$\Phi$ then induces in each
tangent space $T_xM$ a local physical structure $T_x\Phi$,
consisting of
(a) a volume element and
(b) the set of vectors, called causal vectors, which are
tangent to a causal curve through $x$.
Correspondingly, the standard structure $\eta$ in $R^n$
consists of
(a) a volume element
and (b) the image $j^+$ of the set of causal vectors.

About the sets $j^+$ and $j^-:=-j^+$ I make the following
four assumptions.
(i) Except for its direction, the parametrization of a 
causal curve is irrelevant; hence
$j^+$ is scale invariant. 
(ii) But the well-defined direction of causal curves 
distinguishes locally
between past and future; therefore
$j^+\cap j^-=\{0\}$
(`antisymmetry'). 
(iii) As there is no torsion, two causal
vectors at $x\in M$ can be used to construct an
infinitesimal closed geodesic parallelogram;
and as the connection is structure-preserving,
all its four sides must be parts of causal curves.
Requiring transitivity of the causal relation,
its (geodesic) diagonal, too, is part of a causal curve; thus
$j^+$ is convex.
(iv) Let $\theta$ be the coframe field dual
to the $n$-bein.
`No torsion' implies that
for any vector fields $X$, $Y$,
\begin{equation}\label{torsion}
\nabla_X \theta(Y) - \nabla_Y \theta(X) = \theta([X,Y])
\quad.
\end{equation}
Defining $S:={\rm span}\{j^+\}$,
the requirement that the connection be structure-preserving
implies $\nabla_U S
\subseteq S$ for any vector field $U$.
Choosing $X,Y \in \theta^{-1}(S)$ in (\ref{torsion}) one thus obtains
\begin{equation}\label{closed}
[{\theta^{-1}(S)},{\theta^{-1}(S)}] 
\subseteq \theta^{-1}(S)
\quad.
\end{equation}
By Frobenius' theorem,
the causal curves mesh to form a foliation
of $M$, each leaf having dimension at most
${\rm dim}\,S$.
Different leaves would represent `separated worlds',
a situation I want to exclude. Hence ${\rm dim}\,S=n$,
and $j^+$ spans the entire $R^n$.
  
The above properties of the local physical structure
are sufficient to determine the structure group $G$
for $n=2$.
In this case, the boundary
$\partial j^+$ consists of two straight rays.
Choosing basis vectors on these rays,
any symmetry transformation $\Lambda$
has the form $\Lambda = {\rm diag}\,(\lambda,1/\lambda)$
with $\lambda>0$.
The group of these transformations is isomorphic to $SO(1,1)$.
Although for $n\ge 3$ the structure group $G$ is not yet determined,
one may already infer some properties:
as a structure group, $G$ is a Lie group;
it preserves the volume and is therefore
a subgroup of $SL(n,R)$;
it preserves $j^+$ and thus
may not contain reflections:
$(-1)\not\in G$; finally,
\begin{equation}\label{dimg}
{\rm dim}\,G\le n(n-1)/2\quad.
\end{equation}
\textit{Proof}.
{
I prove by induction that any invertible
linear map $g:R^n\to R^n$ preserving $\eta$
is determined by at most $n(n-1)/2$
parameters.
(i) The proposition holds for $n=2$.
(ii) Assume it is proven for $n$.
In $n+1$ dimensions, $g$ is specified by the images
of $n+1$ linearly independent vectors $v_1,\ldots, v_{n+1}$.
These vectors can be chosen 
such that $v_{n+1}$ is the unique
intersection
$v_{n+1}=\partial j^+\cap(v_1+\partial j^-)\cap\ldots
\cap(v_n+\partial j^-)$.
Since $g$ preserves $\partial j^\pm$, the image of
$v_{n+1}$ is uniquely determined by the images of
$v_1,\ldots, v_n$.
It is therefore sufficient to consider the
restriction of $g$ to $S:={\rm span}\{v_1,\ldots,v_n\}$.
And since (by a
suitable choice of the $v_i$) $\partial j^+_s:=S\cap\partial j^+$ can
be made to have
all the properties of a `light cone' in $S$,
it is sufficient to determine the images only
of vectors that lie on $\partial j^+_s$.
To specify $g(\partial j^+_s)=\partial j^+\cap g(S)$
requires at most $n$ parameters; to specify how vectors transform
within $\partial j^+_s$ requires, by assumption, at most
$n(n-1)/2$; so altogether
at most $(n+1)n/2$.
Q.E.D.
}

\section{Deformability}

So far my considerations have been very general,
and the symmetry group $G$ is by no means
uniquely determined. Only now
the key idea of General Relativity comes into play:
rather than being fixed as in Newtonian theory,
the local physical structure on the
space-time manifold is itself a
{\em variable}; it depends on the distribution of matter
in the universe (and on boundary conditions).
Whenever the local physical structure is thus allowed to vary freely
I call the event manifold `deformable'.
However, the `nature' of the physical structure
--- embodied by $\eta$ ---
and hence the symmetry group $G$
must remain unchanged.
Provided the local physical structure reflects
the distribution of matter,
deformability amounts to the requirement that
arbitrary matter distributions be allowed.

Mathematically, varying the $G$-structure corresponds to
varying the $n$-bein; or, equivalently, its dual $\theta$.
By assumption, any choice of the $G$-structure,
and hence of $\theta$, must 
admit a torsion-free $G$-connection.
If expressed with respect to
$\theta$, the connection 1-form $\omega$ takes values in Lie$(G)$
and thus has $n\cdot{\rm dim}\,G$ degrees of freedom.
The requirement that it be torsion-free
reads 
\begin{equation}\label{dbeta}
d\theta+\omega\wedge\theta=0\quad.
\end{equation}
Since $\omega\wedge\theta$ is an $R^n$-valued 2-form,
this requirement imposes $n\cdot n(n-1)/2$ constraints on $\omega$
which for arbitrary $\theta$ can only be satisfied if
$n(n-1)/2\le {\rm dim}\,G$. Together with (\ref{dimg})
one thus obtains
\begin{equation}
{\rm dim}\,G=n(n-1)/2\quad.
\end{equation}
This result implies that the connection
on $M$ is indeed unique,
which Weyl had assumed without proof.
  
  Let us assume that $G$ leaves a proper subspace
$S\subset R^n$ invariant. This again leads to
\ref{closed}, which for
${\rm dim}\,S\ge 2$ is an undue restriction on $\theta$.
Now suppose $S$ is spanned by a vector $r\in R^n$.
The invariance of $S$ implies that there is a 1-form
$\alpha$ such that $\omega r=r\otimes\alpha$.
Defining $X:=\theta^{-1}(r)$ and using (\ref{dbeta})
this yields
\begin{equation}
L_X \theta + \omega(X)\theta = r\otimes\alpha\quad.
\end{equation}
Here $L$ denotes the Lie derivative.
Defining the maps
\begin{equation}
\Lambda : R^n\to R^n \quad,\quad
\Lambda u = -(L_X \theta)(\theta^{-1}(u))
\end{equation}
\begin{equation}
\xi : R^n\to R \quad,\quad
\xi(u) = \alpha(\theta^{-1}(u))
\end{equation}
one obtains
\begin{equation}\label{endo}
\omega(X) - r\otimes\xi = \Lambda\quad,
\end{equation}
a relation among endomorphisms of $R^n$.
Since $\theta$ may be varied freely, $\Lambda$ can be
chosen freely with the sole constraint that
$\Lambda r=0$.
Thus, restricting (\ref{endo}) to 
an $(n-1)$-dimensional 
space $\overline{S}$ complement to $S$ yields
an equation which in general
has a solution only if $\omega(X)|_{\overline{S}}$ and $\xi|_{\overline{S}}$
together have at least as many degrees 
of freedom as $\Lambda|_{\overline{S}}$.
Hence ${\rm dim}\,G +(n-1)\ge n(n-1)$ and therefore
${\rm dim}\,G\ge (n-1)^2$,
a condition which is compatible with 
${\rm dim}\,G=n(n-1)/2$ only if $n=2$.
Thus for $n\ge 3$,
$G$ acts irreducibly on $R^n$.
  
  I have now established several important
properties of the symmetry group $G$ which
are summarized in figure \ref{lofa}.
It can be shown that these 
properties uniquely determine the Lorentz group, 
which in turn implies the existence of an invariant metric
with signature $n-2$.

\section{Conclusion}

{\em Any deformable event manifold is Lorentzian.}
 
This result has a nice physical interpretation if one
assumes a one-to-one correspondence between the manifold's
local physical structure and the distribution of matter:
out of all possible event manifolds, only Lorentzian manifolds
admit arbitrary matter distributions;
any non-metric structure imposes undue restrictions.

\end{document}